\documentclass[twocolumn, superscriptaddress, preview, amsmath, amssymb, aps, letterdraft, notitlepage]{revtex4-1}

\usepackage{graphicx}
\usepackage{amssymb}
\usepackage{color}
\usepackage{booktabs}
\usepackage{subfig}
\usepackage{comment}
\usepackage{appendix}
\usepackage{float}
\usepackage{color}
\definecolor{KB}{rgb}{0.4,0.3,0.9}

\DeclareUnicodeCharacter{0301}{\'{e}}
\begin{document}

\title{Enhancing quantum synchronization through homodyne measurement, noise and squeezing}

\author{Yuan Shen}
\affiliation{School of Electrical and Electronic Engineering, Nanyang Technological University, Block S2.1, 50 Nanyang Avenue, Singapore 639798}
\author{Hong Yi Soh}	
\affiliation{National Institute of Education,
Nanyang Technological University, 1 Nanyang Walk, Singapore 637616}
\author{Weijun Fan}
\email{EWJFan@ntu.edu.sg}
\affiliation{School of Electrical and Electronic Engineering, Nanyang Technological University, Block S2.1, 50 Nanyang Avenue, Singapore 639798}
\author{Leong-Chuan Kwek}
\email{kwekleongchuan@nus.edu.sg}
\affiliation{Centre for Quantum Technologies, National University of Singapore 117543, Singapore}
\affiliation{MajuLab, CNRS-UNS-NUS-NTU International Joint Research Unit, UMI 3654, Singapore}
\affiliation{National Institute of Education,
Nanyang Technological University, 1 Nanyang Walk, Singapore 637616}
\affiliation{School of Electrical and Electronic Engineering, Nanyang Technological University, Block S2.1, 50 Nanyang Avenue, Singapore 639798}

\date{\today}% It is always \today, today,

\begin{abstract}
Quantum synchronization has been a central topic in quantum non-linear dynamics. Despite the rapid development in this field, very few have studied how to efficiently boost synchronization. Homodyne measurement emerges as one of the successful candidates for this task, but preferably in the semi-classical regime. In our work, we focus on the phase synchronization of a harmonic-driven quantum Stuart-Landau oscillator, and show that the enhancement induced by homodyne measurement persists into the quantum regime. Interestingly, optimal two-photon damping rates exist when the oscillator and driving are at resonance and with a small single-photon damping rate. We also report noise-induced enhancement in quantum synchronization when the single-photon damping rate is sufficiently large. Apart from these results, we discover that adding a squeezing Hamiltonian can further boost synchronization, especially in the semi-classical regime. Furthermore, the addition of squeezing causes the optimal two-photon pumping rates to shift and converge.
\end{abstract}

\maketitle

\section{Introduction}
Synchronization pervades nature everywhere, from the unison cacophony of fireflies \cite{winfree1980geometry} to the rhythmic pulse of heart beats \cite{czeisler1986bright} to the locking march steps of superconducting tunnel junctions \cite{zhirov2008synchronization,rosenblum2003synchronization}. We have been captured and mesmerized by the sheer beauty of synchronization. 

Quantum analogues of classical synchronization have been studied in quantum models of self-sustained oscillators \cite{zhirov2006quantum,lee2013quantum, walter2014quantum}. Various measures of quantum synchronization have also been proposed \cite{ameri2015mutual,jaseem2020generalized}. Genuine quantum effects without classical analog, such as synchronization blockade~\cite{lorch2017quantum} and non-linearity-induced synchronization~\cite{shen2023nonlinear_sync}, have also been reported. It has also been shown that additional squeezing can produce stronger frequency entrainment than harmonic drive~\cite{sonar2018squeezing}. Recently, the well-studied field of classical noise-induced synchronization~\cite{zhou2002noise,he2003noise,rozenfeld2001noise,toral2001noise} has also been extended to quantum regime~\cite{schmolke2022}. Experimental works on synchronization, both classical and quantum, have been demonstrated in nanomechanical oscillators \cite{shim2007synchronized,matheny2014phase}, cold atoms\cite{heimonen2018synchronization,laskar2020observation}, quantum dot micropillar\cite{kreinberg2019mutual}, trapped ion qubit~\cite{zhang2022ion_trap_sync} and so forth.

Homodyne measurement is a fundamental technique developed in quantum optics. Yet, it plays a pivotal role in quantum information and technology. Several continuous variable quantum key distribution protocols respond to homodyne detection to extract quadrature information encoded in the signal~\cite{grosshans2002cvqkd,Grosshans2003cvqkd}, which have been demonstrated experimentally~\cite{Jouguet2013,Zhang2019}. It has also been proposed to improve the sensitivity of quantum sensors~\cite{cui2023sensor}. The open quantum system monitored by homodyne measurement can be modeled by a master equation conditioned on the measurement record~\cite{wiseman1993homodyne}, also known as quantum trajectory theory~\cite{wiseman_Quantum_Measurement_and_Control,carmichael2009open}.

In a recent work~\cite{kato2021enhancement}, the phase synchronization of a quantum oscillator with an external drive can be enhanced by monitoring the system with continuous homodyne measurement, and quantum fluctuations are reduced by continuous measurement in the semi-classical regime.  Here we extend the work in Ref.~\cite{kato2021enhancement} to the quantum regime, which leads us to discover more interesting phenomenon in quantum synchronization. We find that the dependence of enhancement on the nonlinear damping rate $\gamma_2$ is not monotonic, and there is an optimal $\gamma_2$ in the semi-classical regime that achieves the greatest enhancement. The most exciting result is the noise-induced synchronization enhancement, where single-photon damping is shown to boost the enhancement. We also report that squeezing can further improve this enhancement as well as make the optimal $\gamma_2$ more robust against noise.

\section{Model}
We study the quantum Stuart--Landau model (also widely accepted as the quantum van~der~Pol model in the existing literature~\cite{walter2014quantum,chia2020relaxation,shen2023nonlinear_sync}) subjected to both a harmonic drive and a two-photon squeeze drive, with continuous homodyne measurement at the output. The stochastic master equation of the system under homodyne measurement in the rotating frame of the drive is given by(with $\hbar = 1$):
\begin{align}
\label{eq:master eq}
    d \rho = &\{-i[\hat{H},\rho] + \gamma_1 \mathcal{D}[a^\dagger]\rho +\gamma_2\mathcal{D}[a^2]\rho \nonumber\\
    &+\gamma_3\mathcal{D}[a]\rho\}dt + \sqrt{\eta_d \gamma_3} \mathcal{H}[a  e^{-i\theta}]\rho dW, \\
    \label{eq:hamiltonian}
    \hat{H} = & -\Delta a^\dagger a + iE(a-a^\dagger) + i \eta(a^{\dagger 2} e^{2i\phi} - a^{2} e^{-2i\phi}), 
\end{align}
where $\mathcal{D}[L]\rho = L\rho L^\dagger - \frac{1}{2}(L^\dagger L \rho + \rho L^\dagger L)$, $\mathcal{H}[L]\rho = L\rho + \rho L^\dagger - \mbox{Tr}[(L+L^\dagger)\rho]\rho$, and $\mathcal{H}[a e^{-i\theta}]$ characterizes the measurement on the quadrature $a e^{-i\theta} + a^\dagger e^{i\theta}$, 
with $\gamma_1$, $\gamma_2$ and $\gamma_3$ corresponding to negative damping, nonlinear damping and linear damping respectively. Without loss of generality we assume a perfect detector and its detection efficiency is set to $\eta_d = 1$ throughout this paper.  We denote $\Delta = \omega_d-\omega_0$ as the amount of initial detuning between the frequency of the drive, $\omega_d$, and the natural frequency of the oscillator, $\omega_0$. $E$ denotes the amplitude of the harmonic drive, with $a$ the annihilation operator and $a^\dagger$ the creation operator. $\eta$\ is the squeezing parameter, and $\phi$ represents the phase of the squeezing. W represents the Wiener process where $\mathbb{E}[dW] = 0$ and $\mathbb{E}[dW^2] = dt$, and the measurement record $dY = \sqrt{\eta_d \gamma_3} \  \mbox{\rm Tr} [(a e^{-i\theta} + a^\dagger e^{i\theta})\rho]dt + dW$.  
We can recover the model described in Ref.~\cite{kato2021enhancement} without feedback control, by setting $\eta=0$. For simplicity, we scale every parameter in the unit of $\gamma_1=1$.

As a measure of synchronization, the phase coherence is frequently used in the literature~\cite{lorch2017quantum,kato2021enhancement}, defined as
\begin{equation}
    S = |S|e^{i\phi_{avg}} = \frac{\mbox{Tr}[a\rho]}{\sqrt{\mbox{Tr}[a^\dagger a\rho]}},
\end{equation}
where $|S|$ measures the degree of phase coherence with a range of $0\le |S|\le 1$. $\phi_{avg}$ represents the average phase of the oscillator. The phase coherence quantifies the statistic fluctuation in the phase distribution and therefore the tendency for quantum oscillator to lock phase with the external drive.

The enhancement of phase coherence through homodyne measurement is calculated as $\mathcal{F}=|S_{HD}|/|S_0|$, where $S_{HD}$ is the average phase coherence over $N_{traj}$ trajectories, defined by
\begin{equation}\label{eq:S_hd}
    S_{HD} = \frac{1}{N_{traj}}\Sigma^{N_{traj}}_{k=1}\frac{\mbox{Tr}[a\rho_k]}{\sqrt{\mbox{Tr}[a^\dagger a\rho_k]}},
\end{equation}
and $S_0$ is the phase coherence obtained from unconditioned master equation without homodyne measurements. We will refer to $\mathcal{F}$ as the enhancement factor. In our simulations, we ensure the finite Hilbert space truncation by examining the fock-space distribution in the steady-states, and we fixed the number of trajectories at $N_{traj}=300$. This allows us to efficiently simulate without utilizing too much computation resources, and further increase the number of trajectories does not change the result qualitatively.

\section{Synchronization enhancement in quantum regime}
\begin{figure}[ht]
    \centering
    \includegraphics[width=\linewidth]{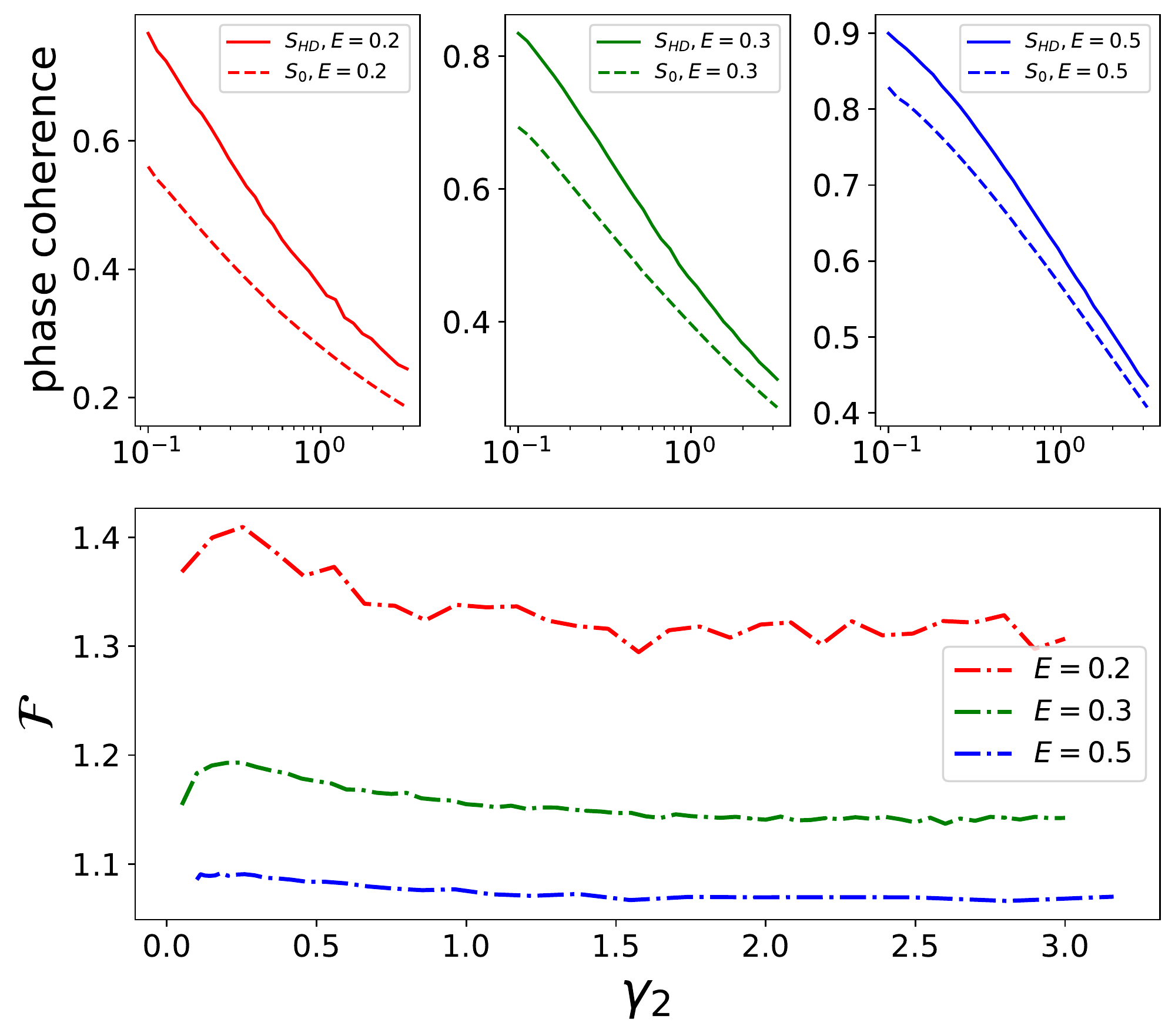}

    \caption{Phase coherence with/without homodyne detection (Top) and enhancement factor $\mathcal{F}$ (Bottom) plotted against nonlinear damping rate $\gamma_2\in [0.5,3]$, under different driving amplitude $E$. Fixed parameters: $\Delta=0,\gamma_3=0.1,\theta=\pi/2$ (optimized for the highest enhancement factor). Take note that the fluctuations in the curves are consequences of a finite number of trajectories averaged in the simulations~(also see Appendix B for simulation errors). The same applies for other plots below.}
    \label{fig:phase coherence}
\end{figure}

We first study the phase synchronization of a quantum Stuart--Landau oscillator with coherent drive but without squeezing Hamiltonian~(by simply set $\eta=0$). Quantum synchronization~(i.e. phase locking) has been shown to improve when the system in semi-classical regime~(the definition of different regimes will be provided later in the section) is continuously monitored by homodyne measurement~\cite{kato2021enhancement}. Here, we investigate the enhancing effect of homodyne measurement in a wider parameter regime and under different driving amplitude. In our simulations, we set zero initial detuning between the oscillator and drive, i.e. $\Delta=0$. This might seem unusual in the context of classical synchronization, where two systems are definitely synchronized without detuning. But here we are interested in the phase synchronization particularly, which is not guaranteed by zero initial detuning. This is due to the presence of quantum fluctuations which prevent the phase-space portrait of the steady-state from concentrating onto one fixed-point, and consequently cause diffusion around it. Another reason for choosing zero initial detuning is that, changing $\gamma_2$ also changes the optimal phase $\theta$ of the measurement when the detuning is non-zero. Whereas with zero detuning, $\theta=\pi/2$ is optimal for all values of $\gamma_2$~(see Appendix A).

\begin{figure}[t!]
    \centering
    \includegraphics[width=\linewidth]{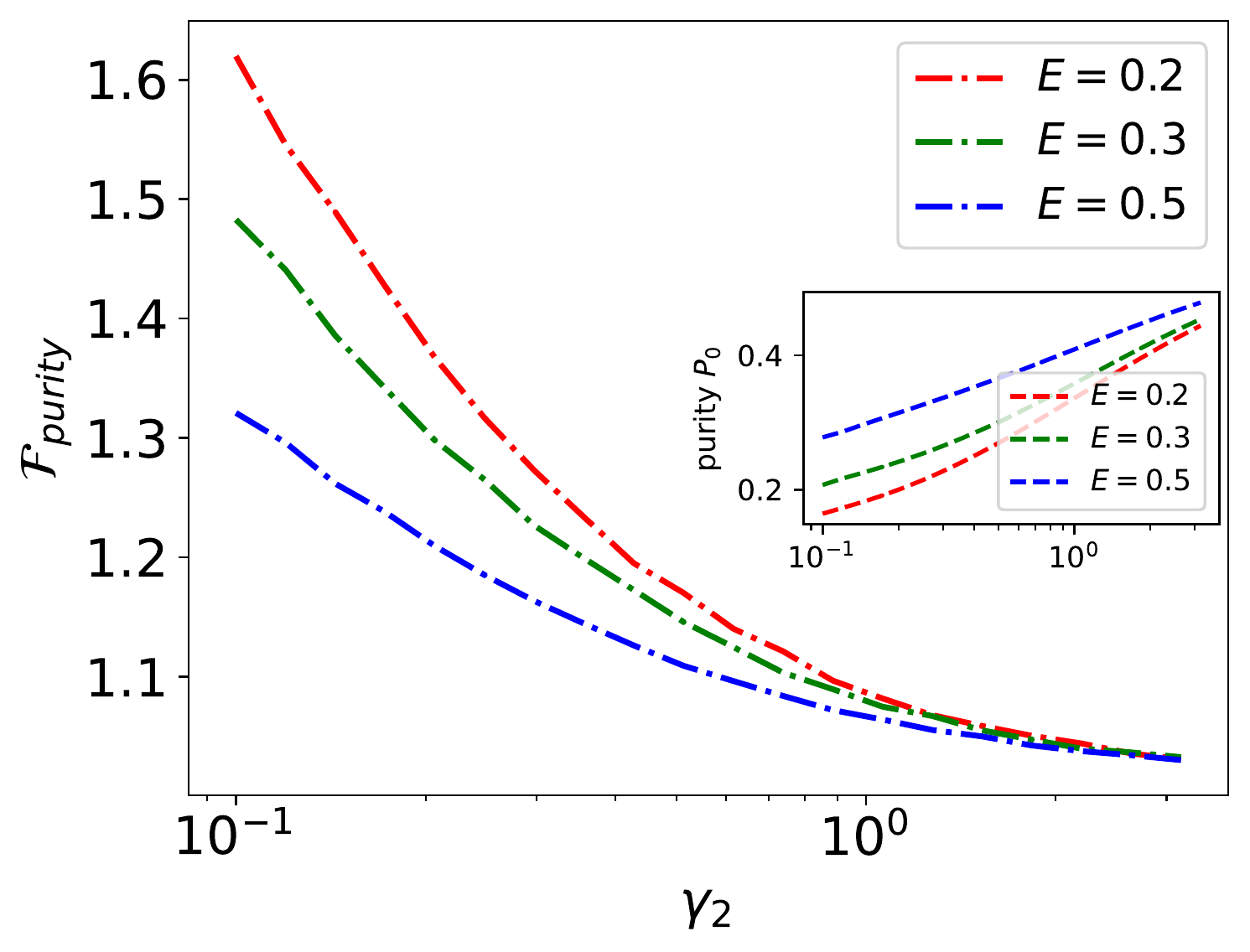}
    \caption{Enhancement in purity of steady-state due to homodyne measurement is always present, but decays with increasing $\gamma_2$. (inset)~Purity scales approximately with $\gamma_2^{1/3}$. Fixed parameters:$\Delta=0,\gamma_3=0.1,\theta=\pi/2$}
    \label{fig:purity}
\end{figure}

In the limit of large mean photon number, the oscillator can be statistically described by a set of classical equation of motion~\cite{walter2014quantum}. This limit is referred to as the semi-classical regime. On the other hand, in the regime with low mean photon number, where the classical model breaks down, the oscillator is considered in the quantum regime. In this regime, quantum noise comes into play and genuine quantum phenomena arise~\cite{dariel2020singlephoton,shen2023nonlinear_sync}. Different synchronization regimes of quantum Stuart--Landau oscillator can be characterized in terms of the nonlinear two-photon damping rate $\gamma_2$. By increasing this rate, the limit cycle of oscillator shrinks and its mean photon number decreases, signifying a transition from semi-classical regime ($\gamma_2<1$) to quantum regime ($\gamma_2\gg1$). In order to access different regimes, we assume the parameter $\gamma_2$ can be tuned freely.
States in quantum regime will be more prone to diffusion through quantum fluctuations, causing them to lose phase synchronization with the driving force. This is shown in the top row panels of Fig.~\ref{fig:phase coherence}, where phase coherence drops with increasing $\gamma_2$, regardless of the driving amplitude. 

In Fig.~\ref{fig:phase coherence} we also show the enhancing effect of homodyne measurement on quantum synchronization, quantified by the enhancement factor $\mathcal{F}$. The presence of homodyne measurement always enhances the phase coherence even in the quantum regime, provided that the phase $\theta$ of the measurement is optimized. 
Larger enhancement is observed when driving amplitude $E$ is small.  Notice in Fig.1, for small $\gamma_2$, there is an optimal ratio for the enhancement factor to peak at, which only appears at zero initial detuning between the oscillator and driving force~(see Appendix A for non-zero detuning cases). After that the enhancement factor drops with $\gamma_2$ until asymptotically reaching a ratio above unity. In addition, later in Section.~\ref{sec:squeezing}, we will show that this optimal ratio is sensitive to the dissipative noise and squeezing.

It has been shown that this enhancement {\cal F} is a consequence
of the increase in the purity of states, defined as $P=\mbox{Tr}[\rho^2]$, as the effective
phase space diffusion is inversely proportional to the purity of states [31]. We note that, on average, homodyne measurement increases
the purity. To this end, we look at the ratio of the purity of the steady state $P_{HD}$ at $\eta_d=1$ , corresponding to homodyne measurement, to the purity $P_0$ at $\eta_d=0$  where homodyne measurement is turned off, i.e. $\mathcal{F}_{purity} = |P_{HD}|/|P_0|$. Fig. 2 shows this ratio with $\gamma_2$. We see that this ratio is always greater than 1, indicating that homodyne measurement enhances the purity of states. However, with increasing $\gamma_2$, this ratio tends to unity, regardless of the amplitude $E$.

\section{Noise-induced synchronization enhancement}

\begin{figure}[t]
    \centering
    \includegraphics[width=\linewidth]{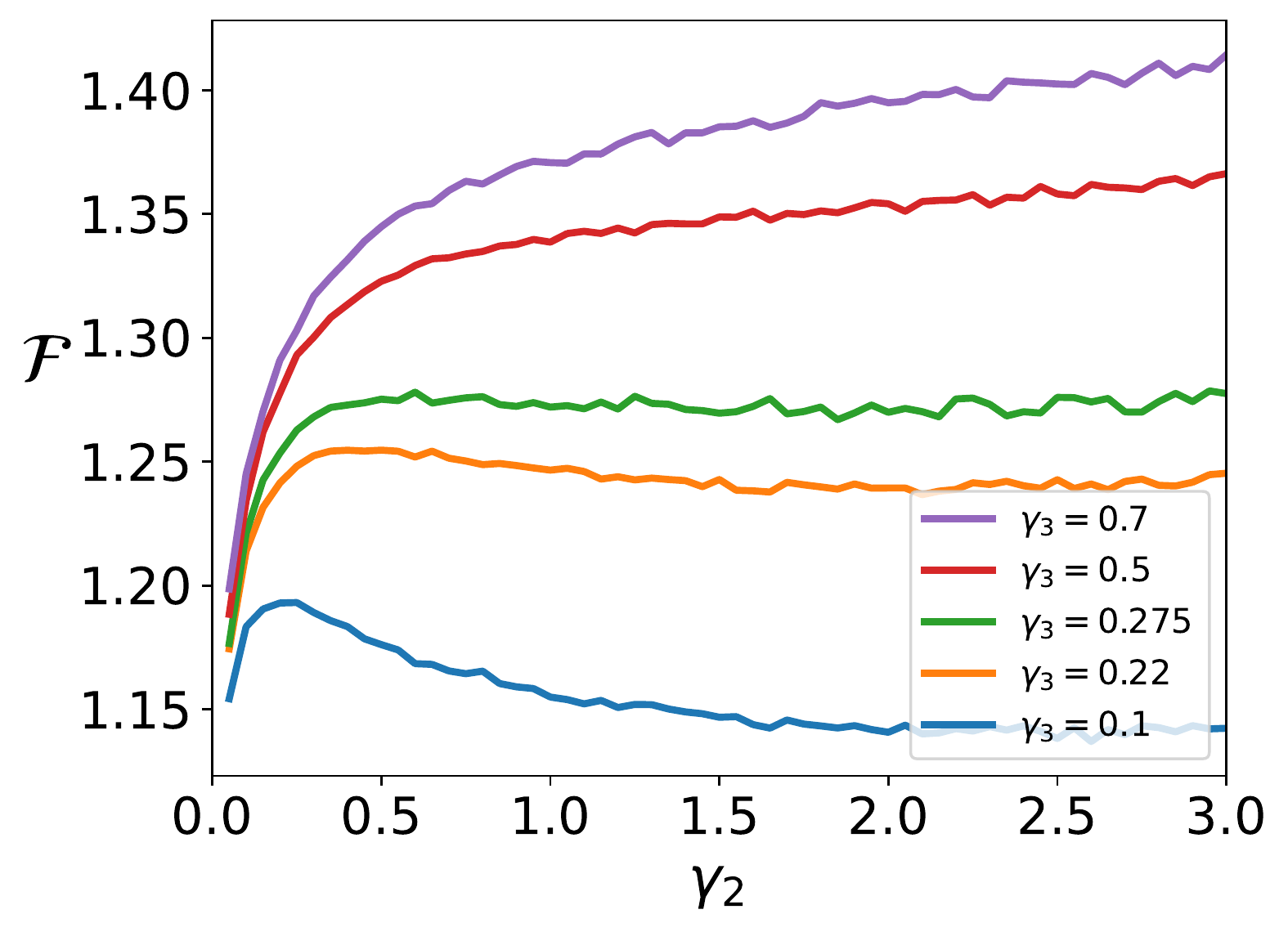}
    \caption{Effect of $\gamma_2$ on enhancement factor $\mathcal{F}$ at different single-photon dissipation $\gamma_3$, with fixed parameters $\Delta=0,E=0.3,\theta=\pi/2$. When single-photon dissipation $\gamma_3$ is small, the enhancement factor $\mathcal{F}$ drops after reaching the maximum. However, the enhancement factor starts to rise even in the quantum regime when a large single-photon dissipation is present. Note that the $\gamma_3=0.1$ curve corresponds to the $E=0.3$ curve in Fig.~\ref{fig:phase coherence}.}
    \label{fig:noise-induced sync}
\end{figure}

\begin{figure}[h]
    \centering
    \includegraphics[width=\linewidth]{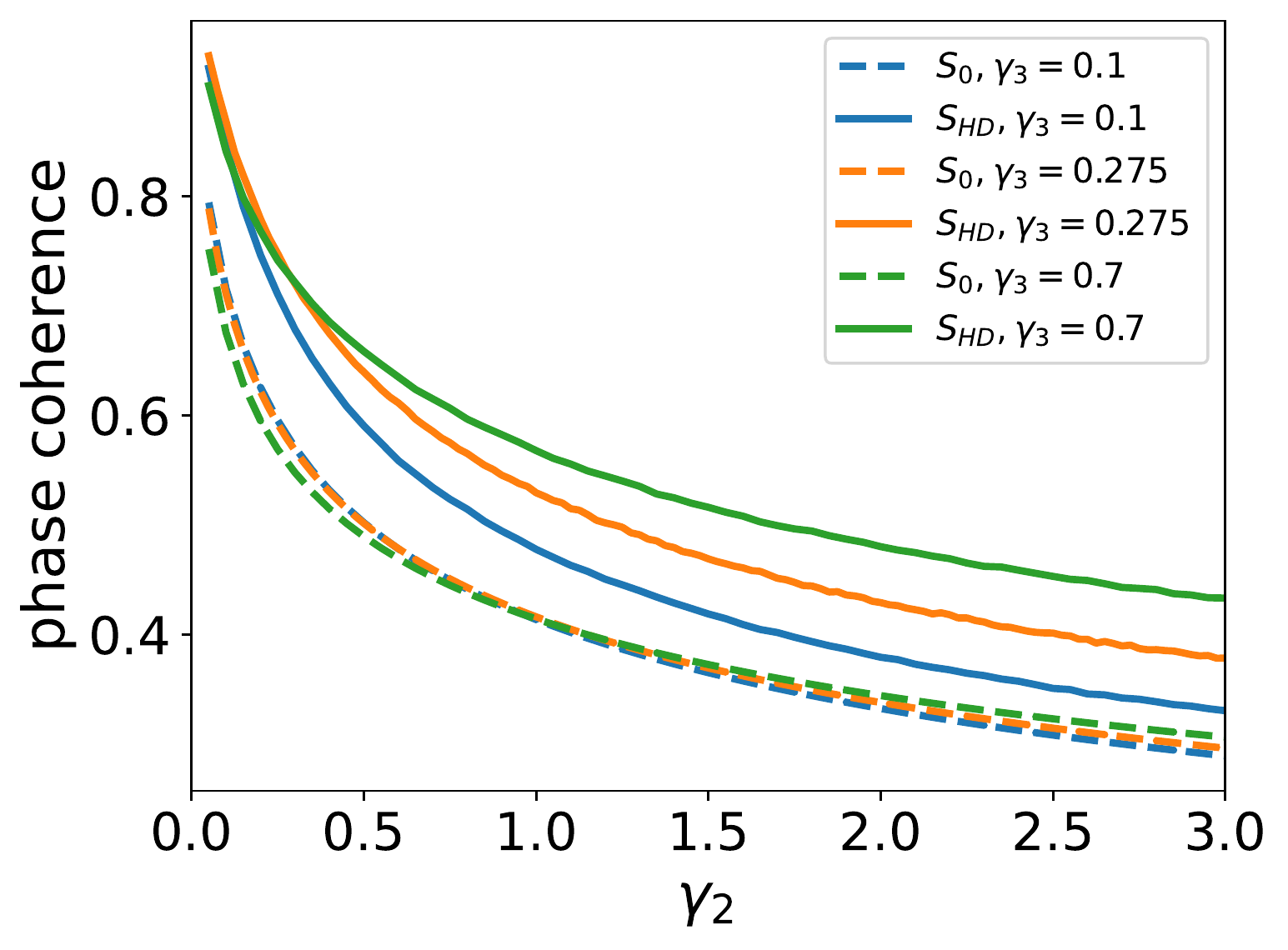}
    \caption{Phase coherence in general vanishes with increasing $\gamma_2$, with fixed parameters $\Delta=0,E=0.3,\theta=\pi/2$. However, in the quantum regime, the presence of single-photon dissipation $\gamma_3$ can boost phase coherence.  Homodyne measurement greatly amplifies this boost in the quantum regime.}
    \label{fig:noise only}
\end{figure}

It has been established that classically chaotic systems exhibit noise-induced synchronization in the presence of common Gaussian noise~\cite{zhou2002noise,toral2001noise}, and this synchronization can be enhanced, for example, by additional dichotomic noise~\cite{rozenfeld2001noise}. In the quantum domain, noise is also reported to boost quantum synchronization, such as inducing frequency entrainment in quantum Stuart--Landau oscillators~\cite{dariel2020singlephoton} and helping to develop correlation and entanglement between two ends of a quantum spin chain~\cite{schmolke2022}. Here we show another example of such noise-induced phase synchronization of a driven quantum Stuart--Landau oscillator.

We show that more single-photon damping can effectively raise the enhancement through homodyne measurement. In Eq.~\eqref{eq:master eq}, $\gamma_3$ corresponds to the rate of single photon damping, acting as a dissipative noise for the oscillator. At the same time, it characterizes the coupling between the measurement device and the monitored system, since all dissipated photons will be captured by the detector~($\eta_d = 1$). The damping rate $\gamma_3$ also scales the back-action of homodyne detection (last term in Eq.~\eqref{eq:master eq}). Therefore, it is not surprising that increasing $\gamma_3$ results in a more potent homodyne measurement and thereby improves the enhancement factor. This is shown in Fig.~\ref{fig:noise-induced sync} where the enhancement factor is plotted against the non-linear damping rate $\gamma_2$ at different single photon damping $\gamma_3$. 
Dissipative noise increases the enhancement factor across all values of $\gamma_2$. Especially in the quantum regime where $\gamma_2 > 1$, the enhancement factor receives a greater boost compared to the semi-classical regime. As a consequence, the enhancement factor increases with non-linear damping rate $\gamma_2$, given a moderate single-photon damping ($\gamma_3 \approx 0.3$) is present. We refer to this as noise-induced synchronization enhancement through homodyne measurement. This is a significant result; compared to Fig.~\ref{fig:phase coherence}, we seem to reverse the effect of increasing $\gamma_2$, from detrimental to enhancing. And this is achieved by simply adding linear damping.

It has also been reported that without any measurements, dissipative noise can produce an increase in off-diagonal density matrix elements in deep quantum regime~\cite{dariel2020singlephoton}, which contributes to greater phase coherence. But such an increase is marginal in amplitude, as seen in our results in Fig.~\ref{fig:noise only}.

\section{Squeezing further enhances synchronization}\label{sec:squeezing}

\begin{figure*}[ht]
    \centering
    \includegraphics[width=\linewidth]{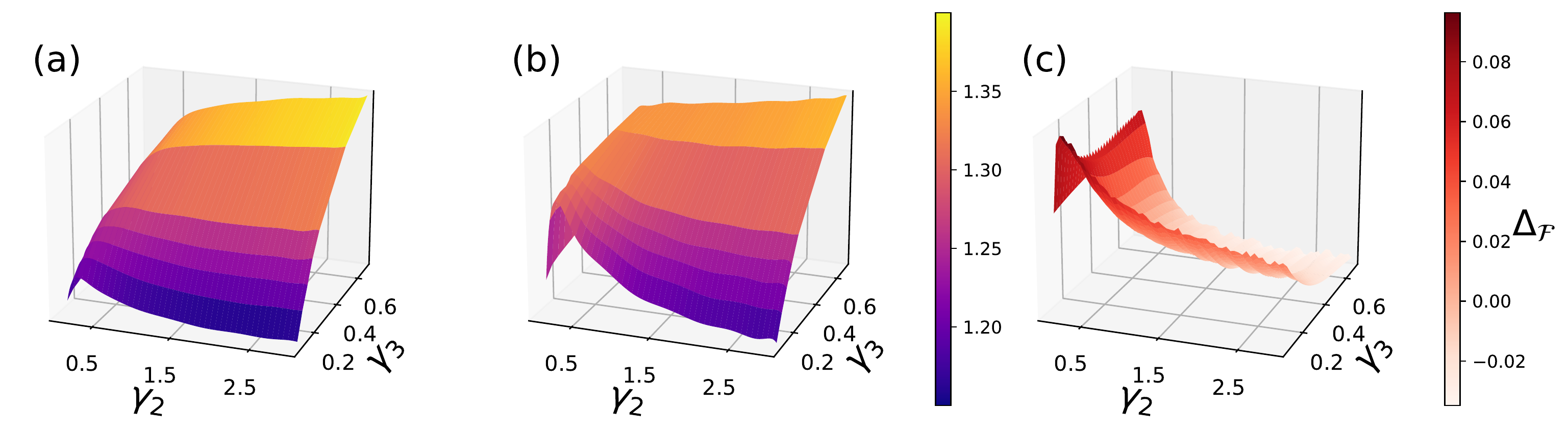}
    \caption{Contour plot of enhancement factor $\mathcal{F}$ against $\gamma_2$ and $\gamma_3$ with (a)$\eta=0$ and (b)$\eta=0.1$; (c) differences between enhancement factor $\mathcal{F}$ of (a) and (b).}
    \label{fig:squeezing}
\end{figure*}

\begin{figure}[ht!]
    \centering
    \subfloat{
    \includegraphics[width=\linewidth]{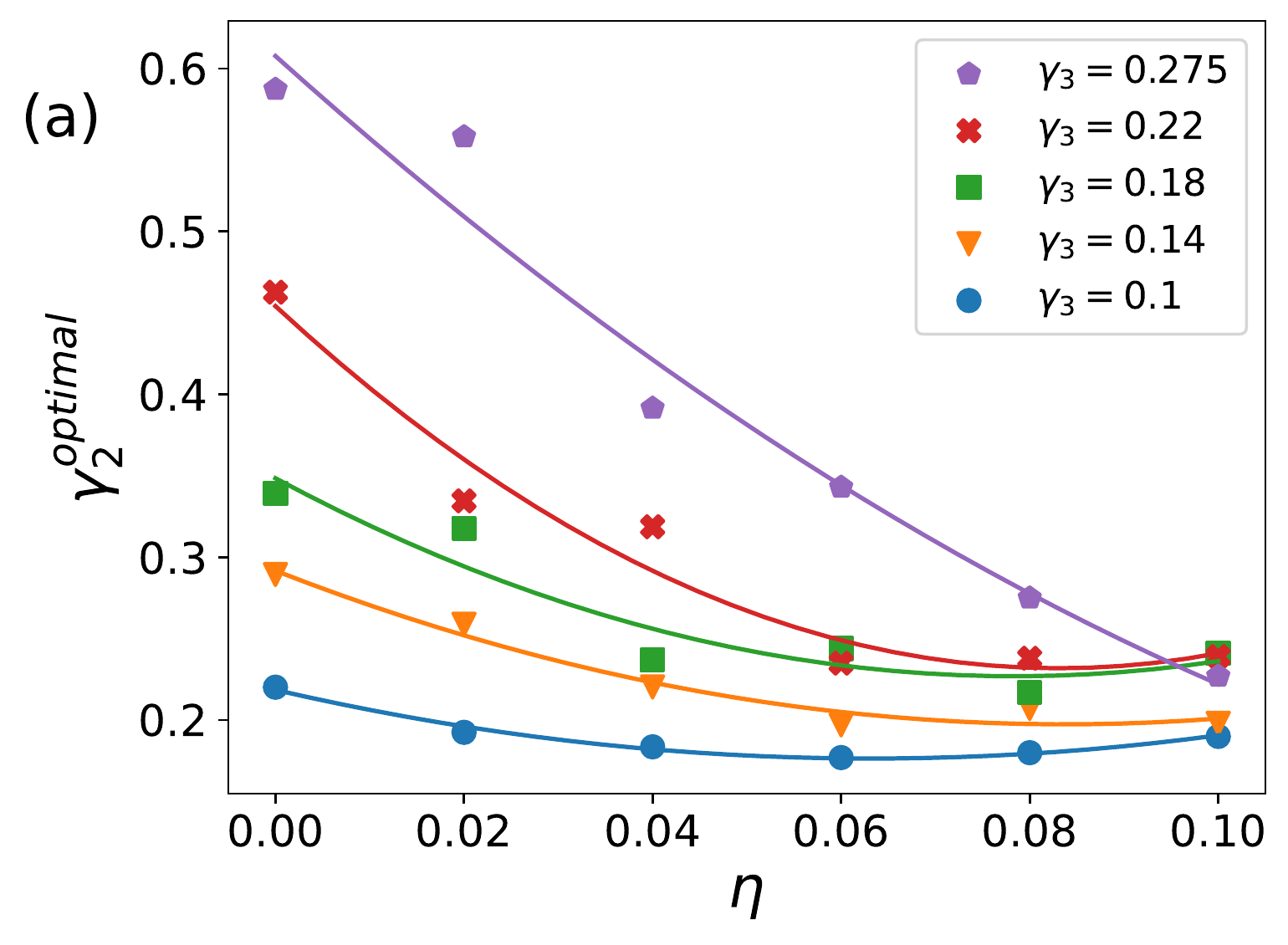}
    }

    \subfloat{
    \includegraphics[width=\linewidth]{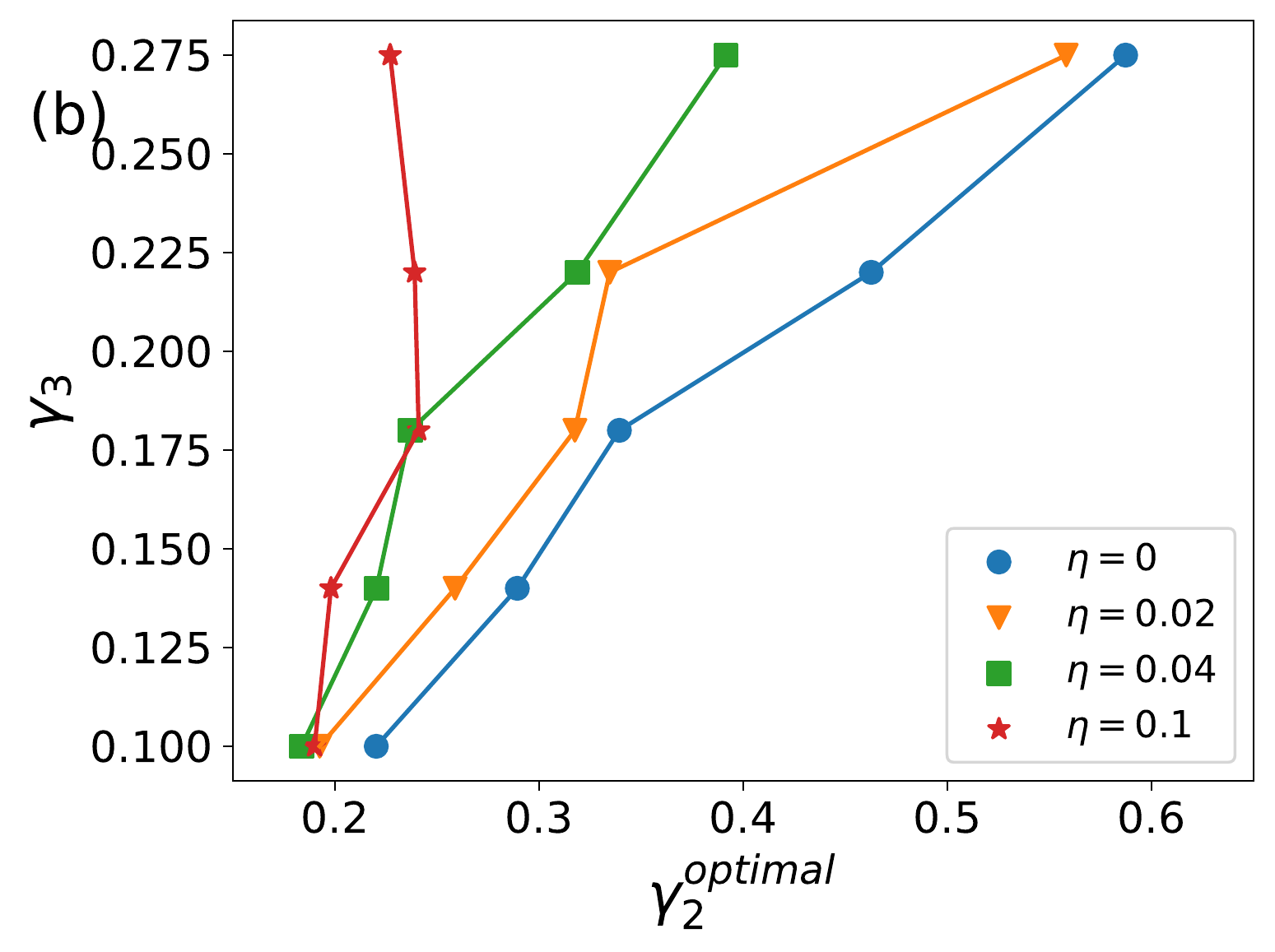}
    }
    \caption{(a)The convergence of optimal $\gamma_2$ with increasing squeezing $\eta$.  (b)Top-down view of the optimal $\gamma_2$ positions in the transition from Fig.5(a)-(b).}
    \label{fig:converging}
\end{figure}

Another useful technique to improve quantum synchronization is squeezing~\cite{sonar2018squeezing}. However, the measure of phase coherence is no longer appropriate for capturing quantum phase synchronization when large squeezing is present. The calculation of phase coherence is valid when the density matrix has only  first off-diagonal coherences~\cite{lorch2017quantum}. Therefore, to study the effects of squeezing, we applied only a small amount of squeezing while ensuring that the second and higher off-diagonal elements remain negligible. It is also known that, under squeezing, the Wigner function of such system will undergo a pitchfork bifurcation~\cite{sonar2018squeezing}, i.e. the steady-state Wigner function has two peaks. In the presence of two peaks, the use of phase coherence as a measure is generally small, due to some form of averaging. Such cases are beyond our scope for this study; therefore, we limit the amplitude of squeezing below $\eta=0.1$ in the $\phi=0$ direction, so that the peak in the Wigner function is not split into two or more peaks. 

Indeed, squeezing is shown to further improve quantum phase synchronization. In Fig.~\ref{fig:squeezing}, we show that squeezing is more likely to boost synchronization in the semi-classical regime, while in the quantum regime, squeezing has negligible effects. The parameter region where squeezing offers the most observable benefits is when both $(\gamma_2,\gamma_3)$ are small~(bottom-left corner of the contour plots). Moving on to the large $(\gamma_2,\gamma_3)$, the enhancing effect of squeezing vanishes, as illustrated in Fig.~\ref{fig:squeezing}(c).  Squeezing and non-linear damping appear to be more closely related to each other, as they are two-photon processes. This is evident from the 3D plots in Fig.~\ref{fig:squeezing}, where increasing squeezing parameter pushes the landscape along $\gamma_2$ rather than $\gamma_3$.

In addition to this additional enhancement caused by squeezing itself, adding squeezing also modifies the enhancement produced by homodyne measurement. Interestingly, the optimal values of $\gamma_2$ for the highest enhancement factor under various $\gamma_3$ converge to  a narrow region around $\gamma_2=0.2$, with a small squeezing up to $\eta=0.1$. This is shown in Fig.~\ref{fig:converging}(a) where we numerically plot the optimal $\gamma_2$ as a function of the squeezing parameter $\eta$. Another interpretation of the effect is that additional squeezing makes the optimal point stronger against dissipative noise $\gamma_3$. As shown in Fig.~\ref{fig:converging}(b), the optimal points $\gamma_2$ move closer to the left vertical axis when the pressure increases, and eventually become independent of $\gamma_3$ when $\eta=0.1$. This effect is interesting as it provides some insights into possible stochastic resonance in the dynamics.

\section{conclusion}
To conclude, our study has made the following observations in quantum synchronization: First, in the presence of homodyne measurement, enhancement in phase synchronization persists to the quantum regime. Next, optimal two-photon nonlinear damping rates ($\gamma_2$) exist in which the enhancement factors are maximum, with small single-photon damping rates ($\gamma_3$) when the oscillator is driven at resonance ($\Delta=0$). This phenomenon is unusual, as it appears only with zero initial detuning. However, moderate single-photon damping rates ($\gamma_3$) allow higher enhancement factors to be achieved even in the quantum regime, despite acting as a source of dissipative noise. Additionally, adding a small amount of squeezing can further enhance quantum phase synchronization, especially in the semi-classical regime. More strikingly, with additional squeeze, optimal non-linear damping rates ($\gamma_2$) become insensitive to dissipative noise ($\gamma_3$).

The Stuart-Landau model used in this work is implementable using the state-of-the-art superconducting circuits, where the two-photon dissipation and squeezing can be engineered using parametric conversion process in Josephson junctions~\cite{dellanno2006multiphoton,leghtas2015confining}. As a future direction, it would be interesting to explore the effects of homodyne measurement and squeezing on a true van~der~Pol oscillator \cite{chia2020relaxation}, where a modified quantum Stuart--Landau oscillator provides a phase space plot that closely resembles the classical diamond-like phase space plot of a van~der~Pol oscillator.

\section*{Acknowledgements}

Numerical simulations are performed using the QuTiP numerical toolbox~\cite{qutip1,qutip2}. YS and WJF would like to acknowledge the support of NRF-CRP19-2017-01 from the National Research Foundation, Singapore. LCK are grateful to the National Research Foundation, Singapore and the Ministry of Education, Singapore for financial support.

\appendix

\section{Enhancement under non-vanishing detuning}

\begin{figure}[ht]
    \centering
    \subfloat[$\Delta=0$]{
    \includegraphics[width=0.9\linewidth]{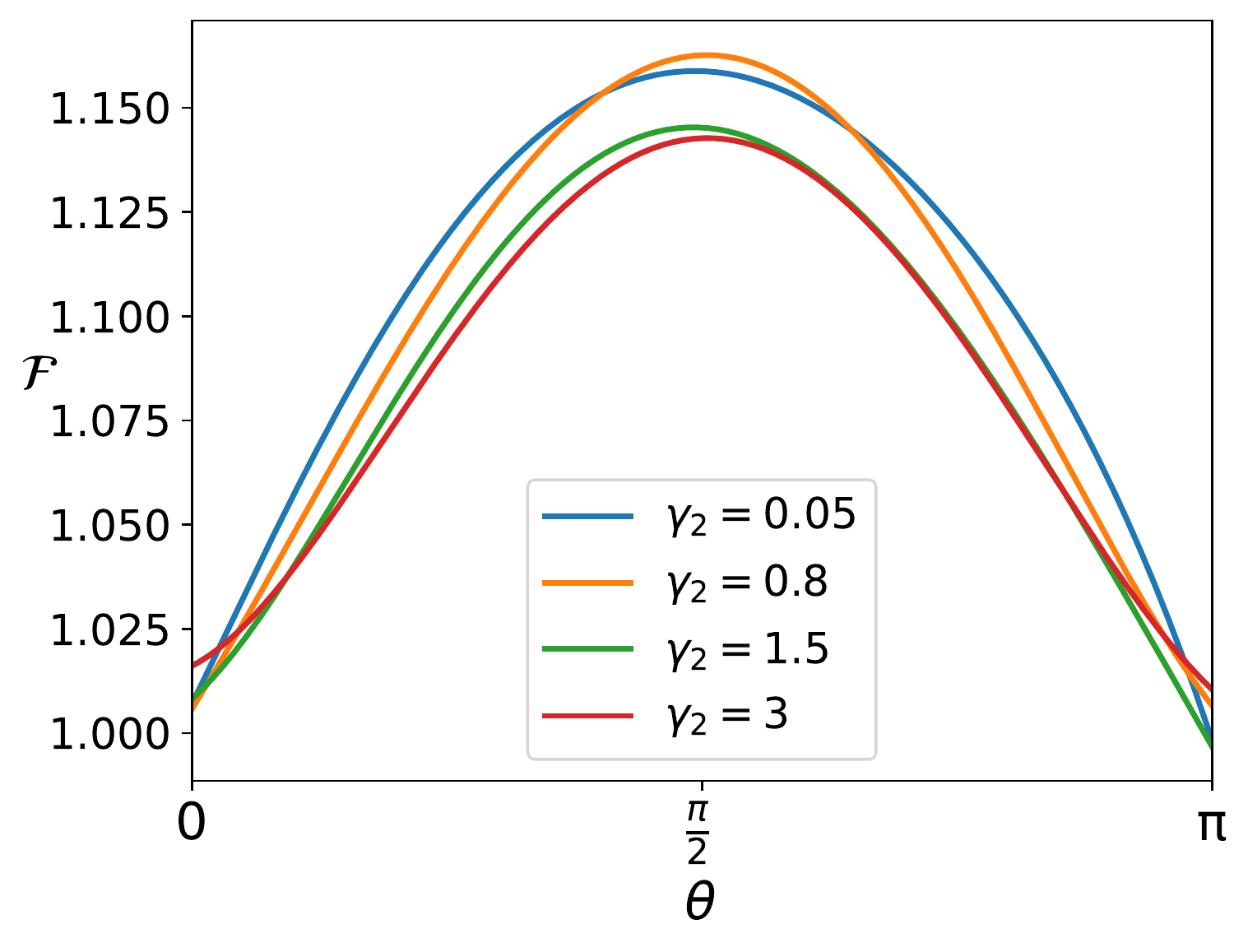}
    }

    \subfloat[$\Delta=0.05$]{
    \includegraphics[width=0.9\linewidth]{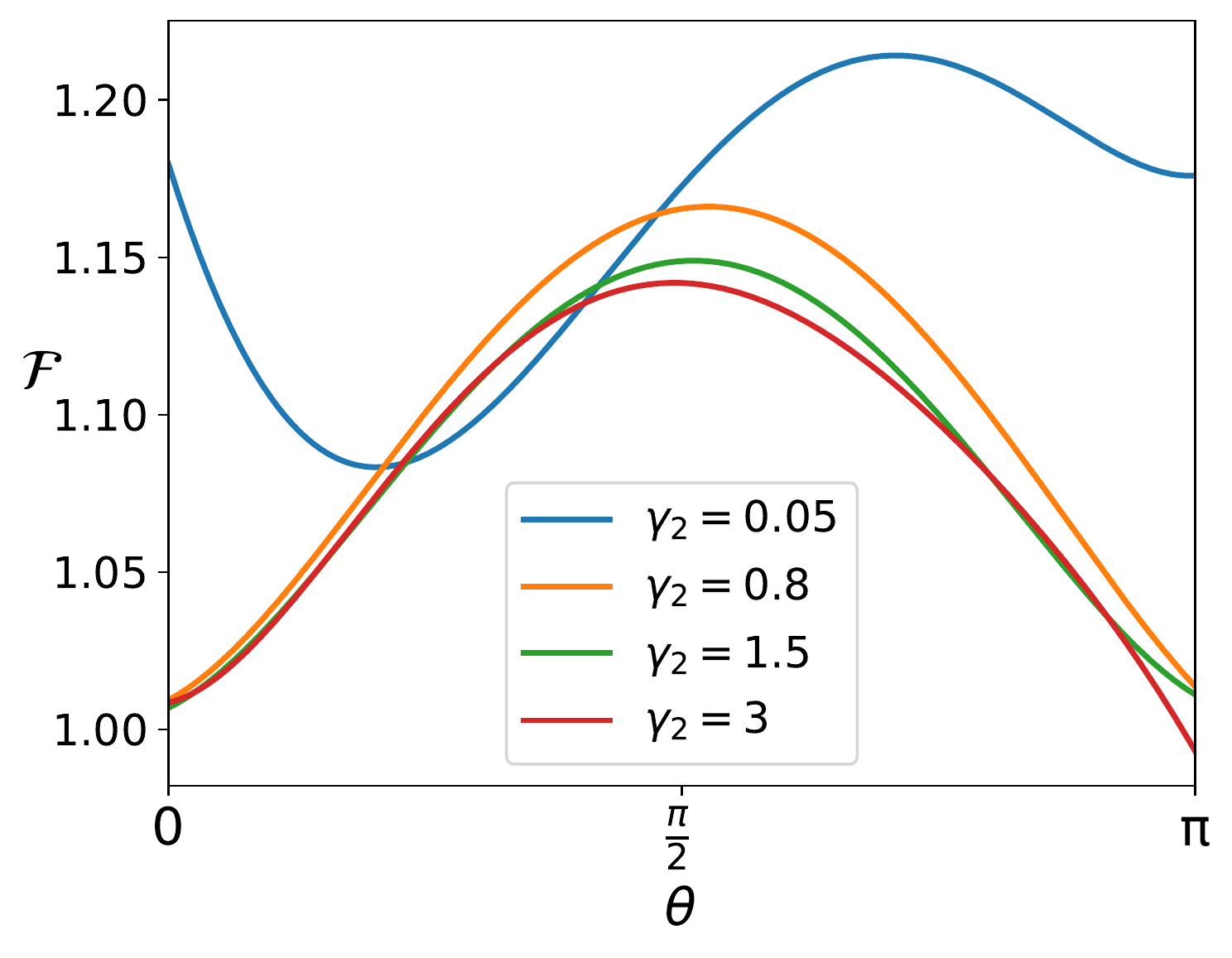}
    }
    \caption{Enhancement factor at different measurement angle $\theta$. (a) Zero detuning. (b) Non-vanishing detuning $\Delta=0.05$. The curves are smoothed using polynomial interpolation.}
    \label{fig:shift_opt_g2}
\end{figure}

In this section, we present the simulation results for non-vanishing detuning, as a supplement reference to the claims we are making. As shown in Fig.~\ref{fig:shift_opt_g2}(a), with zero detuning, the optimal enhancement factors are obtained at $\theta=\pi/2$ and the corresponding measurement angle do not depend on $\gamma_2$. One can also see the optimal enhancement factors do not scale with $\gamma_2$ monotonically, i.e. the highest points of the curves rise and then drop, in the range of $\gamma_2=0.05$ to $3$.

In Fig.~\ref{fig:shift_opt_g2}(b), with non-vanishing detuning, the optimal measurement angles shift away from $\pi/2$. Additionally, the highest enhancement factors are monotonically decreasing with respect to increasing $\gamma_2$.

\section{Simulation error analysis}

\begin{figure}[hbt]
    \centering
    \includegraphics[width=\linewidth]{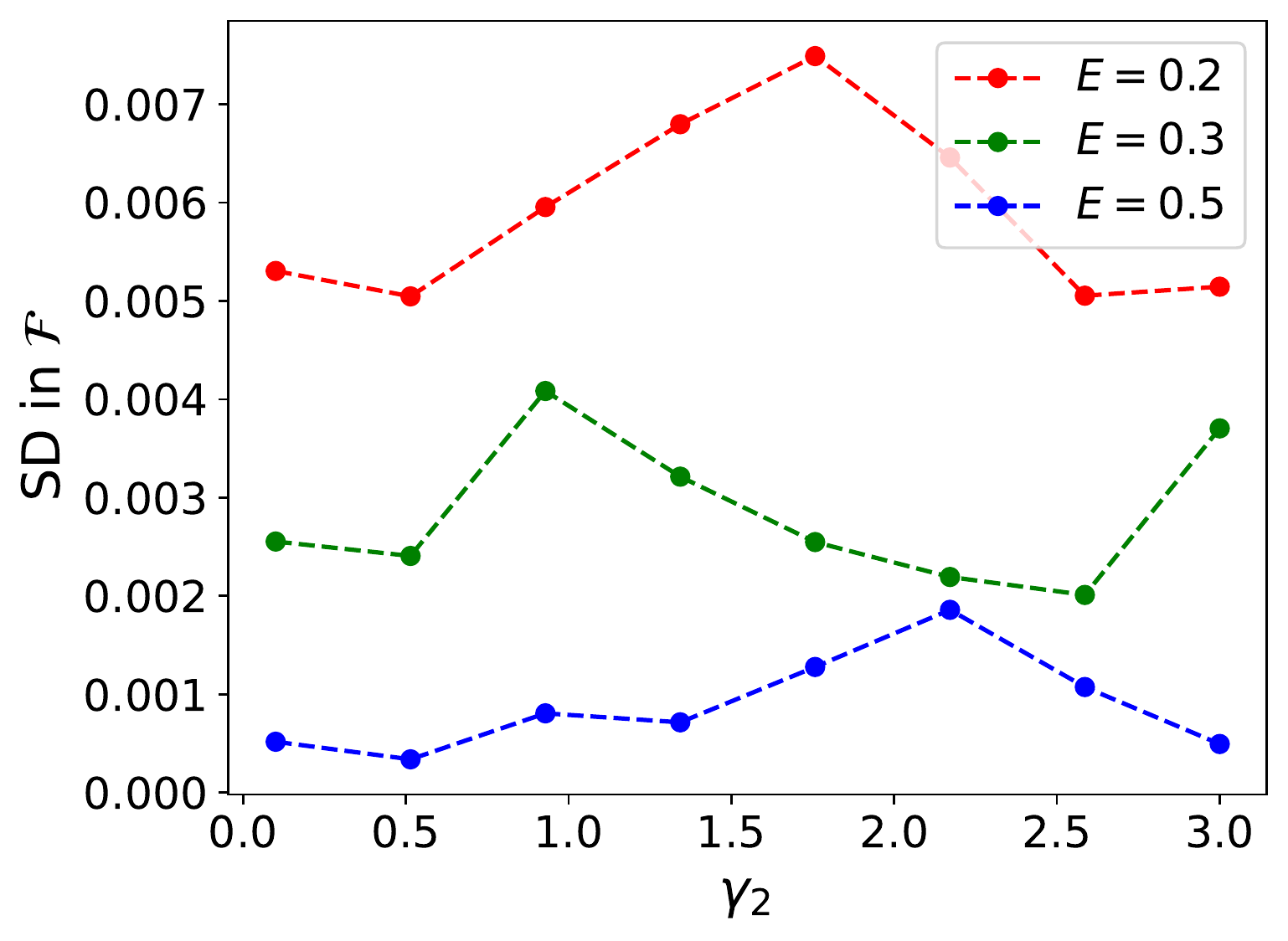}
    \caption{Standard deviation in the enhancement factor $\mathcal{F}$ of Fig.~\ref{fig:phase coherence}.}
    \label{fig:SD}
\end{figure}

We numerically simulated the error (standard deviation) in the data of Fig.~\ref{fig:phase coherence}, with 100 sample runs and number of trajectories $N_{traj}=300$. According to quantum Monte-Carlo simulation method~\cite{ceperley1986qmc}, the standard deviation is on the order of $N_{traj}^{-1/2}. $In Fig.~\ref{fig:SD}, we show that the largest standard deviation is below 1\% of the data in Fig.~\ref{fig:phase coherence}.

\newpage

\bibliographystyle{unsrt} 
\bibliography{ref}

\begin{thebibliography}{10}

\bibitem{winfree1980geometry}
Arthur~T Winfree.
\newblock {\em The geometry of biological time}, volume~2.
\newblock Springer, 1980.

\bibitem{czeisler1986bright}
Charles~A Czeisler, James~S Allan, Steven~H Strogatz, Joseph~M Ronda, Ramiro
  S{\'a}nchez, C~David R{\'\i}os, Walter~O Freitag, Gary~S Richardson, and
  Richard~E Kronauer.
\newblock Bright light resets the human circadian pacemaker independent of the
  timing of the sleep-wake cycle.
\newblock {\em Science}, 233(4764):667--671, 1986.

\bibitem{zhirov2008synchronization}
OV~Zhirov and DL~Shepelyansky.
\newblock Synchronization and bistability of a qubit coupled to a driven
  dissipative oscillator.
\newblock {\em Physical Review Letters}, 100(1):014101, 2008.

\bibitem{rosenblum2003synchronization}
Michael Rosenblum and J{\"u}rgen Kurths.
\newblock {\em Synchronization: a universal concept in nonlinear science}.
\newblock Cambridge University Press, 2003.

\bibitem{zhirov2006quantum}
OV~Zhirov and Dima~L Shepelyansky.
\newblock Quantum synchronization.
\newblock {\em The European Physical Journal D-Atomic, Molecular, Optical and
  Plasma Physics}, 38(2):375--379, 2006.

\bibitem{lee2013quantum}
Tony~E Lee and HR~Sadeghpour.
\newblock Quantum synchronization of quantum van der pol oscillators with
  trapped ions.
\newblock {\em Physical Review Letters}, 111(23):234101, 2013.

\bibitem{walter2014quantum}
Stefan Walter, Andreas Nunnenkamp, and Christoph Bruder.
\newblock Quantum synchronization of a driven self-sustained oscillator.
\newblock {\em Physical Review Letters}, 112(9):094102, 2014.

\bibitem{ameri2015mutual}
V~Ameri, M~Eghbali-Arani, A~Mari, A~Farace, F~Kheirandish, V~Giovannetti, and
  R~Fazio.
\newblock Mutual information as an order parameter for quantum synchronization.
\newblock {\em Physical Review A}, 91(1):012301, 2015.

\bibitem{jaseem2020generalized}
Noufal Jaseem, Michal Hajdu{\v{s}}ek, Parvinder Solanki, Leong-Chuan Kwek,
  Rosario Fazio, and Sai Vinjanampathy.
\newblock Generalized measure of quantum synchronization.
\newblock {\em Physical Review Research}, 2(4):043287, 2020.

\bibitem{lorch2017quantum}
Niels L{\"o}rch, Simon~E Nigg, Andreas Nunnenkamp, Rakesh~P Tiwari, and
  Christoph Bruder.
\newblock Quantum synchronization blockade: Energy quantization hinders
  synchronization of identical oscillators.
\newblock {\em Physical Review Letters}, 118(24):243602, 2017.

\bibitem{shen2023nonlinear_sync}
Yuan Shen, Wai-Keong Mok, Changsuk Noh, Ai~Qun Liu, Leong-Chuan Kwek, Weijun
  Fan, and Andy Chia.
\newblock Quantum synchronization effects induced by strong nonlinearities.
\newblock {\em Phys. Rev. A}, 107:053713, May 2023.

\bibitem{sonar2018squeezing}
Sameer Sonar, Michal Hajdu{\v{s}}ek, Manas Mukherjee, Rosario Fazio, Vlatko
  Vedral, Sai Vinjanampathy, and Leong-Chuan Kwek.
\newblock Squeezing enhances quantum synchronization.
\newblock {\em Physical Review Letters}, 120(16):163601, 2018.

\bibitem{zhou2002noise}
Changsong Zhou and J\"urgen Kurths.
\newblock Noise-induced phase synchronization and synchronization transitions
  in chaotic oscillators.
\newblock {\em Phys. Rev. Lett.}, 88:230602, May 2002.

\bibitem{he2003noise}
Daihai He, Pengliang Shi, and Lewi Stone.
\newblock Noise-induced synchronization in realistic models.
\newblock {\em Phys. Rev. E}, 67:027201, Feb 2003.

\bibitem{rozenfeld2001noise}
Robert Rozenfeld, Jan~A. Freund, Alexander Neiman, and Lutz Schimansky-Geier.
\newblock Noise-induced phase synchronization enhanced by dichotomic noise.
\newblock {\em Phys. Rev. E}, 64:051107, Oct 2001.

\bibitem{toral2001noise}
Raúl Toral, Claudio~R. Mirasso, Emilio Hernández-Garcı́a, and Oreste Piro.
\newblock Analytical and numerical studies of noise-induced synchronization of
  chaotic systems.
\newblock {\em Chaos: An Interdisciplinary Journal of Nonlinear Science},
  11(3):665--673, 2001.

\bibitem{schmolke2022}
Finn Schmolke and Eric Lutz.
\newblock Noise-induced quantum synchronization.
\newblock {\em Phys. Rev. Lett.}, 129:250601, Dec 2022.

\bibitem{shim2007synchronized}
Seung-Bo Shim, Matthias Imboden, and Pritiraj Mohanty.
\newblock Synchronized oscillation in coupled nanomechanical oscillators.
\newblock {\em Science}, 316(5821):95--99, 2007.

\bibitem{matheny2014phase}
Matthew~H Matheny, Matt Grau, Luis~G Villanueva, Rassul~B Karabalin, MC~Cross,
  and Michael~L Roukes.
\newblock Phase synchronization of two anharmonic nanomechanical oscillators.
\newblock {\em Physical Review Letters}, 112(1):014101, 2014.

\bibitem{heimonen2018synchronization}
Hermanni Heimonen, Leong~Chuan Kwek, Robin Kaiser, and Guillaume Labeyrie.
\newblock Synchronization of a self-sustained cold-atom oscillator.
\newblock {\em Physical Review A}, 97(4):043406, 2018.

\bibitem{laskar2020observation}
Arif~Warsi Laskar, Pratik Adhikary, Suprodip Mondal, Parag Katiyar, Sai
  Vinjanampathy, and Saikat Ghosh.
\newblock Observation of quantum phase synchronization in spin-1 atoms.
\newblock {\em Physical Review Letters}, 125(1):013601, 2020.

\bibitem{kreinberg2019mutual}
S{\"o}ren Kreinberg, Xavier Porte, David Schicke, Benjamin Lingnau, Christian
  Schneider, Sven H{\"o}fling, Ido Kanter, Kathy L{\"u}dge, and Stephan
  Reitzenstein.
\newblock Mutual coupling and synchronization of optically coupled quantum-dot
  micropillar lasers at ultra-low light levels.
\newblock {\em Nature communications}, 10(1):1--11, 2019.

\bibitem{zhang2022ion_trap_sync}
Liyun Zhang, Zhao Wang, Yucheng Wang, Junhua Zhang, Zhigang Wu, Jianwen Jie,
  and Yao Lu.
\newblock Observing quantum synchronization of a single trapped-ion qubit,
  2022.

\bibitem{grosshans2002cvqkd}
Fr\'ed\'eric Grosshans and Philippe Grangier.
\newblock Continuous variable quantum cryptography using coherent states.
\newblock {\em Phys. Rev. Lett.}, 88:057902, Jan 2002.

\bibitem{Grosshans2003cvqkd}
Fr{\'e}d{\'e}ric Grosshans, Gilles Van~Assche, J{\'e}r{\^o}me Wenger, Rosa
  Brouri, Nicolas~J. Cerf, and Philippe Grangier.
\newblock Quantum key distribution using gaussian-modulated coherent states.
\newblock {\em Nature}, 421(6920):238--241, Jan 2003.

\bibitem{Jouguet2013}
Paul Jouguet, S{\'e}bastien Kunz-Jacques, Anthony Leverrier, Philippe Grangier,
  and Eleni Diamanti.
\newblock Experimental demonstration of long-distance continuous-variable
  quantum key distribution.
\newblock {\em Nature Photonics}, 7(5):378--381, May 2013.

\bibitem{Zhang2019}
G.~Zhang, J.~Y. Haw, H.~Cai, F.~Xu, S.~M. Assad, J.~F. Fitzsimons, X.~Zhou,
  Y.~Zhang, S.~Yu, J.~Wu, W.~Ser, L.~C. Kwek, and A.~Q. Liu.
\newblock An integrated silicon photonic chip platform for continuous-variable
  quantum key distribution.
\newblock {\em Nature Photonics}, 13(12):839--842, Dec 2019.

\bibitem{cui2023sensor}
Dianzhen Cui, Jianning Li, Fude Li, Zhi-Cheng Shi, and X.~X. Yi.
\newblock Enhancing the sensitivity of nonlinearity sensors through homodyne
  detection in dissipatively coupled systems.
\newblock {\em Phys. Rev. A}, 107:013709, Jan 2023.

\bibitem{wiseman1993homodyne}
H.~M. Wiseman and G.~J. Milburn.
\newblock Quantum theory of field-quadrature measurements.
\newblock {\em Phys. Rev. A}, 47:642--662, Jan 1993.

\bibitem{wiseman_Quantum_Measurement_and_Control}
Howard~M. Wiseman and Gerard~J. Milburn.
\newblock {\em Quantum Measurement and Control}.
\newblock Cambridge University Press, 2009.

\bibitem{carmichael2009open}
Howard Carmichael.
\newblock {\em An open systems approach to quantum optics: lectures presented
  at the Universit{\'e} Libre de Bruxelles, October 28 to November 4, 1991},
  volume~18.
\newblock Springer Science \& Business Media, 2009.

\bibitem{kato2021enhancement}
Yuzuru Kato and Hiroya Nakao.
\newblock Enhancement of quantum synchronization via continuous measurement and
  feedback control.
\newblock {\em New Journal of Physics}, 23(1):013007, 2021.

\bibitem{chia2020relaxation}
Andy Chia, Leong~Chuan Kwek, and C~Noh.
\newblock Relaxation oscillations and frequency entrainment in quantum
  mechanics.
\newblock {\em Physical Review E}, 102(4):042213, 2020.

\bibitem{dariel2020singlephoton}
W.-K. Mok, L.-C. Kwek, and H.~Heimonen.
\newblock Synchronization boost with single-photon dissipation in the deep
  quantum regime.
\newblock {\em Phys. Rev. Research}, 2:033422, Sep 2020.

\bibitem{dellanno2006multiphoton}
Fabio Dell’Anno, Silvio {De Siena}, and Fabrizio Illuminati.
\newblock Multiphoton quantum optics and quantum state engineering.
\newblock {\em Physics Reports}, 428(2):53--168, 2006.

\bibitem{leghtas2015confining}
Z.~Leghtas, S.~Touzard, I.~M. Pop, A.~Kou, B.~Vlastakis, A.~Petrenko, K.~M.
  Sliwa, A.~Narla, S.~Shankar, M.~J. Hatridge, M.~Reagor, L.~Frunzio, R.~J.
  Schoelkopf, M.~Mirrahimi, and M.~H. Devoret.
\newblock Confining the state of light to a quantum manifold by engineered
  two-photon loss.
\newblock {\em Science}, 347(6224):853--857, 2015.

\bibitem{qutip1}
J.R. Johansson, P.D. Nation, and Franco Nori.
\newblock Qutip: An open-source python framework for the dynamics of open
  quantum systems.
\newblock {\em Computer Physics Communications}, 183(8):1760--1772, 2012.

\bibitem{qutip2}
J.R. Johansson, P.D. Nation, and Franco Nori.
\newblock Qutip 2: A python framework for the dynamics of open quantum systems.
\newblock {\em Computer Physics Communications}, 184(4):1234--1240, 2013.

\bibitem{ceperley1986qmc}
David Ceperley and Berni Alder.
\newblock Quantum monte carlo.
\newblock {\em Science}, 231(4738):555--560, 1986.

\end{thebibliography}

\end{document}